\documentclass[iop]{emulateapj}
\begin{document}
\newcommand{\lnsigt}{\ln{10}\hspace{2pt}\sigma_T}
\title{Quasar Accretion Disks Are Strongly Inhomogeneous}
\shorttitle{Inhomogeneous Disks}
\shortauthors{Dexter \& Agol}
\author{Jason Dexter}
\affil{Department of Physics, University of Washington, Seattle, WA 98195-1560, USA}
\email{jdexter@u.washington.edu}
\author{Eric Agol}
\affil{Department of Astronomy, University of Washington, Box 351580, Seattle, WA 98195, USA}
\keywords{accretion, accretion disks --- black hole physics --- galaxies: active --- gravitational lensing: micro}
\begin{abstract}
Active galactic nuclei (AGN) have been observed to vary stochastically with $10-20\%$ rms amplitudes over a range of optical wavelengths where the emission arises in an accretion disk. Since the accretion disk is unlikely to vary coherently, local fluctuations may be significantly larger than the global rms variability. We investigate toy models of quasar accretion disks consisting of a number of regions, $n$, whose temperatures vary independently with an amplitude of $\sigma_T$ in dex. Models with large fluctuations ($\sigma_T=0.35-0.50$) in $10^{2-3}$ independently fluctuating zones for every factor of two in radius can explain the observed discrepancy between thin accretion disk sizes inferred from microlensing events and optical luminosity while matching the observed optical variability. For the same range of $\sigma_T$, inhomogeneous disk spectra provide excellent fits to the \emph{HST} quasar composite without invoking global Compton scattering atmospheres to explain the high levels of observed UV emission. Simulated microlensing light curves for the Einstein cross from our time-varying toy models are well fit using a time-steady power-law temperature disk, and produce magnification light curves that are consistent with current microlensing observations. Deviations due to the inhomogeneous, time-dependent disk structure should occur above the $1\%$ level in the light curves, detectable in future microlensing observations with millimag sensitivity.
\end{abstract}

\maketitle

\begin{figure*}
\epsscale{1.1}
\plotone{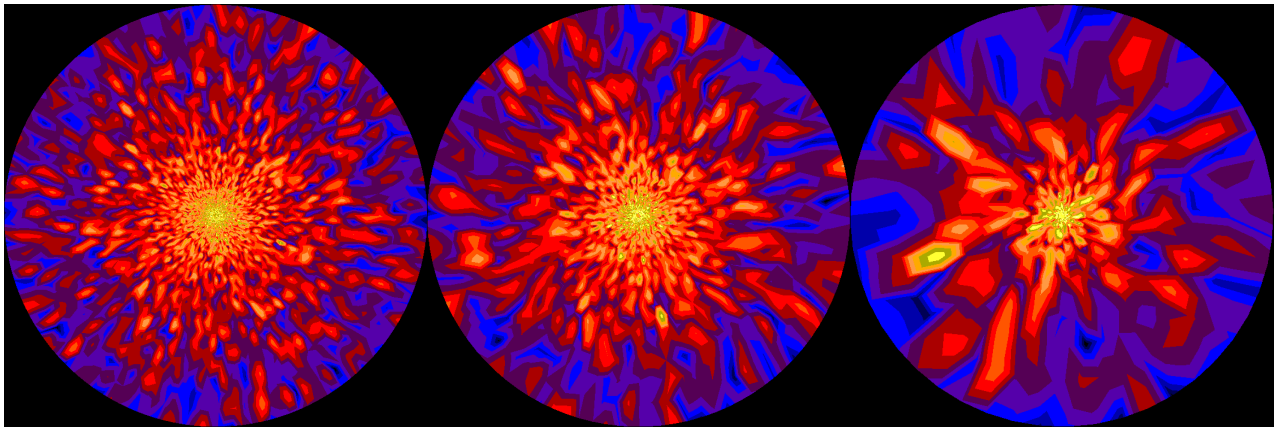}
\caption{\label{timgs}False color temperature maps of damped random walk models with $n=2200$ (left), $n=550$ (middle) and $n=140$ (right). The images are interpolated onto a Cartesian grid for plotting, and the color scale increases logarithmically from blue to red to yellow to white with a dynamic range of 100.}
\end{figure*}

\section{Introduction}

Optical emission from active galactic nuclei (AGN) is thought to be due to thermal emission from a standard thin accretion disk \citep{shaksun1973,novthorne}. This model is successful in explaining many observations of black hole accretion disks, such as the peak frequencies of the thermal spectra of X-ray binaries and AGN. It is also used to fit continuum spectra in the X-ray to measure black hole spin \citep[e.g.,][]{shafee2006}, and to infer accretion disk size from gravitational microlensing; both in absolute terms \citep[e.g.,][]{morganetal2010} and as a function of wavelength \citep{anguitaetal2008,poindexteretal2008,eigenbrodetal2008,morganetal2010}. Attempts to fit continuum AGN spectra with thin disk models have met with mixed success \citep{blaesetal2001,davisetal2007}. While multi-temperature blackbody spectra generally fit well in the optical, at UV wavelengths quasar spectra are brighter than predicted \citep[e.g.,][]{blaesetal2001}. The X-ray size has been shown by microlensing to be extremely compact \citep{chartasetal2009,daietal2010}, disfavoring spectral models invoking global Comptonizing atmospheres to explain the UV emission.

Quasar optical variability is also difficult to explain in the context of the thin disk model. Long term monitoring of AGN has found almost simultaneous variability across optical wavelengths, with lags of less than 1-2 days \citep{cutrietal1985,claveletal1991,koristaetal1995,wandersetal1997,collieretal1998}. Comparing these lags to the radii dominating the thin disk emission at these wavelengths gives a traveling speed of 0.1c for the variability mechanism \citep{kroliketal1991,courvoisierclavel1991}. This would force the variability in AGN to be communicated at the local sound or Alfv\'{e}n speed rather than on the infall timescale associated with disk instabilities.

\citet{wambsganssetal1990} and \citet{rauchblandford1991} first measured the size of an accretion disk from  microlensing, and found that accretion disks were smaller than expected from the thin disk model for the observed $L_\nu$. In recent years, the opposite trend has emerged \citep{pooleyetal2007,daietal2010,morganetal2010,blackburneetal2010}. Sizes are now robustly found to be a factor of $\sim4$ larger on average than expected from the thin disk model at IR/optical/UV wavelengths. The size vs. wavelength relation predicted by the thin disk, $r \propto \lambda^{4/3}$, is within the large range allowed by microlensing observations \citep{eigenbrodetal2008}. At the same time, \citet{kellyetal2009}, \citet{kozlowskietal2010} and \citet{macleod2010} have studied large samples of quasar light curves. They find that optical quasar variability is well described by a damped random walk which returns to a mean value on a typical timescale of 200 days with variability amplitudes of $\simeq 10-20\%$.

It is unlikely that the observed quasar variability is caused by a coherently varying accretion disk\footnote{The implications of coherent variations for microlensing measurements are discussed by \citet{blackburnekochanek2010}.}, but rather is probably the added effect of many smaller, independently varying regions. Many such models have been proposed to explain quasar variability \citep[e.g.][]{lyubarskii1997}. In this Letter, we demonstrate that for the observed variability characteristics, such an inhomogeneous disk can simultaneously explain multiple discrepancies between AGN observations and accretion disk theory. Inhomogeneous disks can be large enough to explain the microlensing observations, while their temperature fluctuations on small spatial scales naturally explain the observed simultaneous variability across optical wavelengths. Temperatures exceeding the local thin disk value lead to broader spectra extending into the UV, consistent with quasar spectra without invoking a Compton scattering medium.

\begin{figure}
\epsscale{1.2}
\plotone{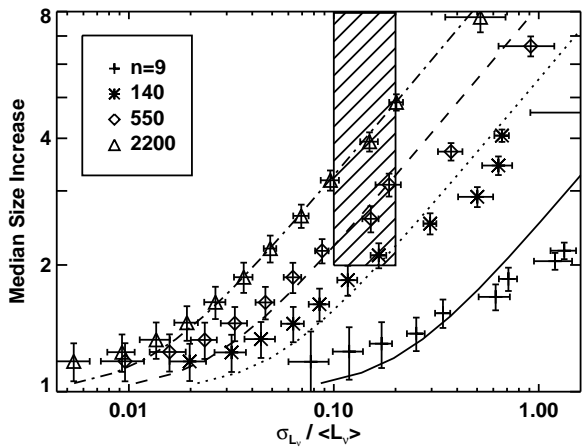}
\caption{\label{rarrsdmn}Median size increase vs. relative variability for various zone sizes for the damped random walk model. Each sequence has values of $\sigma_T$ from $0.1-0.8$. The allowed region from microlensing and variability studies is shaded. The lines show analytic fits to the variability from the damped random walk model (Eq. \ref{varfit}), and to the fractional increase in half-light radius from the log-normal model (Eq. \ref{sizefit}). For $n \gtrsim 100$ and $\sigma_T \simeq 0.4$, damped random walk disks can be large enough to explain microlensing while matching the observed variability.}
\end{figure}

\section{Inhomogeneous Accretion Disks}

We assume that i) the optical/UV emission observed from AGN originates in an optically thick accretion disk. ii) Variations in the disk occur locally, and are uncorrelated on large spatial scales. iii) Fluid in the disk is on circular Keplerian orbits. Assumption i) allows us to model the disk emission using temperature alone, while ii) requires that the disk be inhomogeneous. Assumption iii) is only used for computing microlensing light curves, where the disk surface brightness map enters into the magnification light curve produced.

\citet{kellyetal2009} found that quasar light curves are well described as a CAR(1) process, a random walk that tends to return to a mean value on a typical timescale. The timescale they found was roughly 200 days, consistent with the thermal timescale in the predicted optical emission region of many AGN. This behavior was confirmed for SDSS Stripe 82 quasars by \citet{macleod2010}. The amplitude of variations in typical sources is $10-20\%$. To produce this variability amplitude with multiple, independent regions in the disk requires larger local variations (total variance $\propto N^{-1}$ for $N$ independent zones). The local accretion disk flux and
effective temperature will then no longer be a monotonic
function of radius $r$, but will fluctuate with azimuth and
time.  This will cause the disk spectrum within an annulus
to have stronger emission at shorter wavelengths than if
the same flux were emitted at constant temperature.
Consequently, the outer portions of the disk will contribute
more flux than for a uniform disk, causing the disk to appear
larger at a particular wavelength.

The thin disk model\footnote{In this Letter, we use ``thin disk model'' to mean that of \citet{shaksun1973}. Inhomogeneous disks may also be geometrically thin.} is broadly consistent with many observed properties of black hole accretion flows, and follows from the conservation of angular momentum and energy in the gravitational potential of the black hole. Its global time-averaged properties are likely correct. Assuming a geometrically thin, optically thick disk the local flux can be written $F_\nu (r,\phi,t)=\pi B_\nu (T)$, where $F_\nu$ is the flux at the frequency $\nu$, $B_\nu$ is the Planck function and $T=(F/\sigma_b)^{1/4}$, and $\sigma_b$ is the Stefan-Boltzmann constant. In the thin disk case, $T \propto r^{-3/4}$ well outside the inner disk edge and is independent of $\phi$ and $t$ everywhere. 

\begin{figure}
\epsscale{1.2}
\plotone{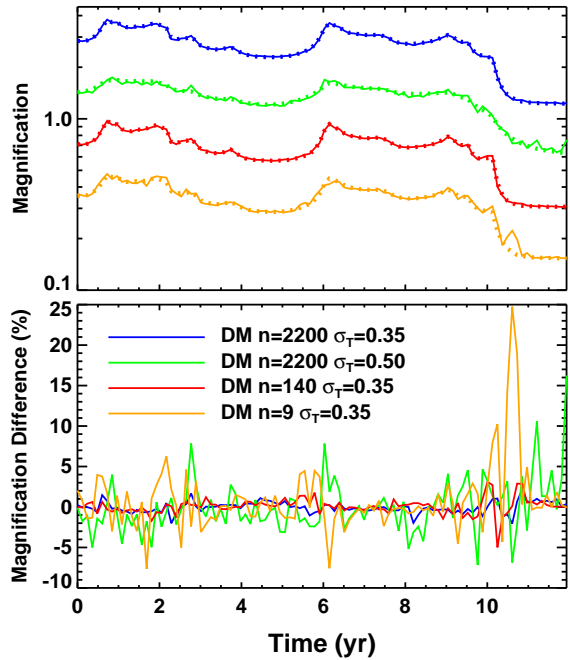}
\caption{\label{mlens}Top: Sample microlensing light curves for all models, shifted for legibility. The mean of each curve is $\simeq1.3$. Best fit power-law disk model light curves are shown as dotted lines. Bottom: Fractional difference between inhomogeneous and best fit power-law disk magnification light curves. Deviations are largest at small $n$ and large $\sigma_T$.}
\end{figure}

\section{Damped Random Walk Disk Model}

We allow $T$ to be a function of $\phi$ and $t$ as well as $r$ by dividing the disk into $n$ evenly spaced zones in $\log{r}$ and $\phi$ per octave in radius. Motivated by variability studies, we let $\log_{10} T$ follow an independent damped random walk in each zone with an amplitude $\sigma_T$ that is independent of radius. The local temperature returns to a mean value on a characteristic timescale of 200 days that is also taken to be independent of radius, and the mean value is chosen so that $\langle \sigma_b T^4 \rangle_{\phi,t} = F(r)$, where $F(r)$ is the thin disk flux at the radius $r$. The results in this paper are completely insensitive to the choice of characteristic timescale and its radial dependence. The total flux is calculated by integrating over the area of the disk assuming face-on viewing and ignoring all relativistic effects. The accretion flow undergoes Keplerian rotation, but the zones remain completely independent at all times. Sample temperature maps from this model for a few values of $n$ are shown in Figure \ref{timgs}.

Figure \ref{rarrsdmn} shows the ratio of half-light radii between the damped random walk disk and a standard thin disk at the same flux as a function of the variability, $\sigma_{L_\nu}$. The overall variability is fit well by the expression,

\begin{equation}
\label{varfit}
 \sigma_{L_\nu}^2 \simeq \frac{3}{8 n} (e^{3\sigma_T^2}-1) \langle L_\nu \rangle^2.
 \end{equation} 
 
For a large number of zones, the disk can be significantly larger than a thin disk while producing little variability. For few (large) zones, the overall variability becomes unrealistically large before the accretion disk becomes large enough to explain the microlensing measurements. By design, power spectra and structure functions from this damped random walk model are consistent with studies of quasar variability \citep{kellyetal2009,kozlowskietal2010,macleod2010}. This model is by no means unique. We have also tried other time-variable toy models for inhomogeneous disks, which can reproduce the observed variability characteristics. All models considered give similar size-variability amplitude relations. 

In the limit $n \rightarrow \infty$, the damped random walk model becomes time-independent with a log-normal distribution of disk temperatures in each annulus whose variance is $(\lnsigt)^2/2$. In this limit, the flux from an annulus can be written,

\begin{equation}
  F_\nu (r;\sigma_T) = \frac{2\pi h \nu^3}{c^2} \int_0^\infty \frac{dw}{\sqrt{\pi }\lnsigt w} \frac{e^\frac{-[\ln{w}+(\lnsigt)^2]^2}{(\lnsigt)^2}}{e^{z/w}-1},
\end{equation}

\noindent where $z \equiv h\nu/kT$, $T$ is the thin disk temperature, and the local temperature in an infinitesimal piece of the annulus is $w T$. The specific luminosity can be computed from,

\begin{equation}
  L_\nu (\sigma_T) = \int_0^\infty 2\pi r F_\nu(r;\sigma_T) dr.
\end{equation}

The size increase as a function of $\sigma_T$ is the ratio of half-light radii from this model assuming $T \propto r^{-3/4}$ to those computed from $L_\nu (0)$. Sizes are compared at fixed $L_\nu$ since the half-light radius, $r_h$, is a function of $L_\nu$ ($r_h \propto \sqrt{L_\nu}$ ignoring the disk inner edge). The relation, 

\begin{equation}
\label{sizefit}
   r_h(\sigma_T)/r_h(0)=e^{0.85 (\lnsigt)^2},
\end{equation} 

\noindent gives an excellent fit to numerical calculations. In models with $\sigma_T=0.35-0.5$, $5-10\%$ of the zones in a given annulus produce $50\%$ of its flux in agreement with the analytical log-normal ($n \rightarrow \infty$) model. The fractional area contributing half the flux decreases with increasing $n$ and $\sigma_T$.

\subsection{Microlensing Simulations}

Thus far, we have compared ratios of inhomogeneous to smooth model thin disk half-light radii at the same value of $L_\nu$. However, microlensing observations do not measure half-light radii. Instead, they measure uncorrelated variability between different images of strongly lensed quasars, caused by the motion of stars in the lens galaxy. Sizes can be measured by assuming a power-law thin disk with $T(r) \propto r^{-3/4}$ and fitting magnification light curves with surface brightness maps parameterized by the radius, $r_s$, where $h\nu=kT(r_s)$ \citep{kochanek2004}. For the power-law disk, $r_s$ can also be calculated from the observed $L_\nu$ in the absence of all lensing effects. The discrepancy between the values of $r_s$ inferred from microlensing versus $r_s$ inferred from $L_\nu$ is $0.6 \pm 0.3$ dex \citep{morganetal2010}. In the power-law disk, $r_h = 2.44 r_s$. We would like to determine whether microlensing light curves from our inhomogeneous toy model can be fit well by the power-law disk model, and whether the values of $r_s$ inferred from microlensing simulations of the inhomogeneous model are consistent with our previous results for the half-light radius. 

Using the code described by \citet{wambsganss1999}, we simulate magnification patterns for QSO 2237+0305, the Einstein Cross, using lens galaxy parameters found by \citet{kochanek2004}. Choosing random orientations and starting positions, many magnification light curves are produced from the time-varying toy models at multiple observed frequencies, and for the power-law disk at many values of $r_s$. The microlensing light curves are then fit to determine best fit values of $r_s$. Sample light curves are shown in the top panel of Figure \ref{mlens} along with best fit power-law disk light curves, while the bottom panel shows fractional deviations between the two.

The time-steady, power-law disk model produces excellent fits to the microlensing light curves. In the damped random walk model, $r_s$ from microlensing is consistent with the half light radius of the disk at various frequencies by a fixed fraction that depends on $n$. For many (few) zones, $r_s$ underestimates (overestimates) the time-averaged disk half-light radius. The largest deviations are of order $30\%$ and will not dominate the error in estimating quasar accretion disk sizes from microlensing. 

\begin{figure}
\epsscale{1.2}
\plotone{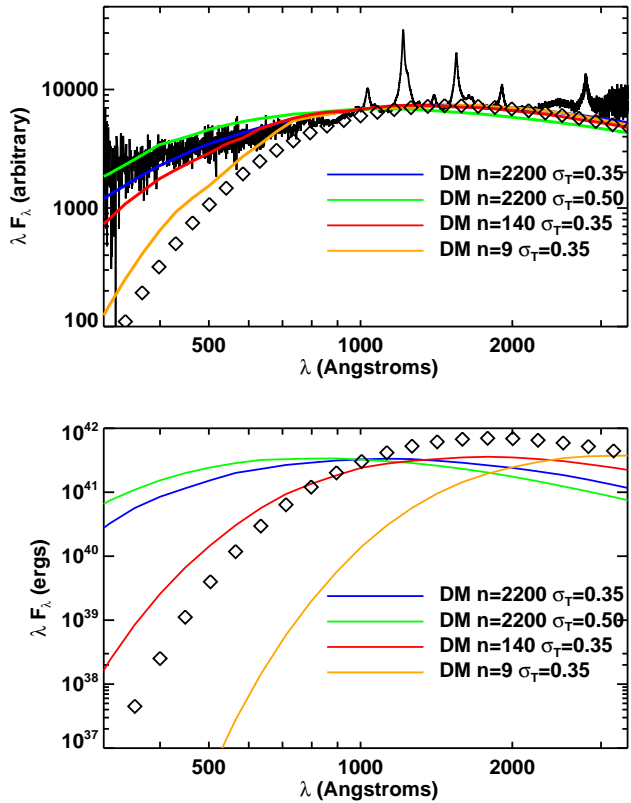}
\caption{\label{compspec}Top panel: Model spectra compared to the composite from \citet{zhengetal1997}. The open diamonds show a thin disk spectrum. Bottom panel: Model spectra (lines) compared to thin disk spectrum (diamonds) for an annulus at $r=20r_{\mathrm{ms}}$, where $r_{\mathrm{ms}}$ is the inner disk edge. The black hole mass is $10^8 M_\sun$ with $L/L_{edd}=0.1$.}
\end{figure}

\begin{figure}
\epsscale{1.2}
\plotone{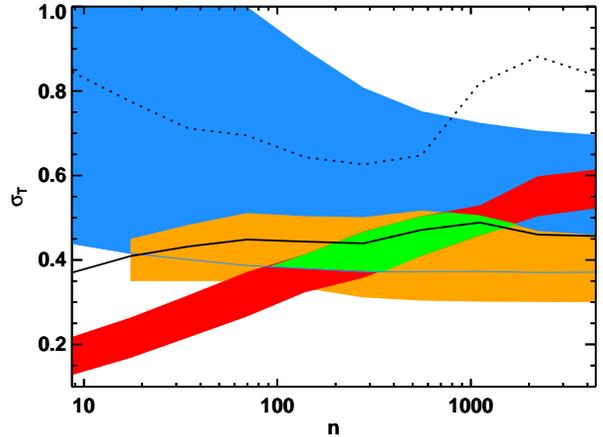}
\caption{\label{constraints}Allowed $68\%$ regions of the $\sigma_T$ vs. $n$ parameter space from microlensing (\emph{blue}), variability (\emph{red}) and spectral (\emph{orange}) observations. Their intersection is shaded \emph{green}. Contours of $1\%$ (\emph{solid}) and $2\%$ (\emph{dotted}) rms deviation
between model microlensing light curves and best fit power-law
temperature disks are overplotted.}
\end{figure}

\subsection{Spectra and Light Curves}

Damped random walk model spectra are compared to the composite quasar spectrum from \citet{zhengetal1997} in the top panel of Figure \ref{compspec}. The temperature normalizations have been chosen to provide the best fit. In all cases, the toy model spectra have larger $F_\lambda$ at smaller $\lambda$ than the thin accretion disk model. The bottom panel of Figure \ref{compspec} shows local spectra at $r=20r_{\mathrm{ms}}$, where $r_{\mathrm{ms}}$ is the inner disk edge. The local inhomogeneous spectra peak at a smaller value of $\lambda F_\lambda$ than the single temperature blackbody. To produce the same total flux at that wavelength, the emission must arise from a larger area. Thus, the disk appears larger at that wavelength.

These spectral differences will cause parameter values inferred from optical spectra to differ between the two models. We find that inferred black hole masses (accretion rates) are smaller (larger) when fitting inhomogeneous spectra with thin disk models. The discrepancy is typically less than a factor of two, and is unlikely to be a significant source of error in estimating accretion disk parameters from spectral fitting.

The damped random walk model produces simultaneous optical variability, in that the cross-correlation between all wavebands peaks at a lag of zero timesteps. The lag is zero regardless of the chosen timestep, because neighboring zones are completely independent. Variability is stronger at higher frequencies, in qualitative agreement with observations. Quantitative predictions for this relation, for cross-correlations between light curves at different observed wavelengths, or for trends with luminosity or black hole mass will require a physical model for the variability mechanism. These quantities are highly sensitive to the particular inhomogeneous disk prescription chosen, many of which satisfy the existing observational constraints.

\subsection{Combined Observational Constraints}

Figure \ref{constraints} summarizes our main results. The $68\%$ allowed regions of the $\sigma_T$ vs. $n$ parameter space from microlensing, spectral and variability measurements are shown for the damped random walk model, along with contours of rms deviations between model and best fit power-law disk microlensing light curves at $2500$\AA. The disk size and variability constraints are the same as in Figure \ref{rarrsdmn}. The spectral constraint is found from fitting model spectra to the composite spectrum from \citet{zhengetal1997}, varying the temperature normalization in each case to obtain the best fit.

The independent spectral and microlensing size constraints require similar amplitudes of temperature inhomogeneity. Producing the observed level of variability requires larger $\sigma_T$ at larger $n$. All constraints intersect in the green region, giving best fit parameter values of $0.35 \lesssim \sigma_T \lesssim 0.5$ and $100 \lesssim n \lesssim 1500$. The models predict short timescale deviations from smooth power-law disk models in microlensing light curves at or above the $1\%$ level on average, with maximum deviations of $10\%$ (see Figure \ref{mlens}). These are likely to be lower limits. Deviations from other toy models tried can be a factor of a few larger.

\section{Physical Mechanisms}

Quasar accretion disks are expected to be inhomogeneous to some degree. However, to produce the observed sizes and broad spectra observed in AGN, the inhomogeneities must be large -- factor of $3-5$ variations in temperature at a given radius. Disk instabilities are the most promising mechanism for producing inhomogeneous disks. The magnetorotational instability \citep[MRI, ][]{mri} is now believed to operate in a wide range of accretion flows, causing the disk to become fully turbulent and providing a means for outward angular momentum transport and accretion. Local \citep[e.g.,][]{millerstone2000} and global simulations of thin MRI disks have been performed both in pseudo-Newtonian potentials \citep{armitageetal2001,armitagereynolds2003} and recently in full general relativity \citep{shafee2008,noble2009,noblekrolik2010,pennaetal2010}. None of these simulations included radiation forces, which likely dominate the dynamics of AGN accretion disks over a large range in radius \citep{shaksun1973}. \citet{hiroseetal2009} studied a set of local simulations in radiative MHD, many of which were radiation-dominated. Their simulations show temperature variations with factors of $\sim2$ in a local patch of disk. Global general relativistic MHD simulations from \citet{fragile2007} and \citet{mckinneyblandford2009} have gas temperatures that can vary by factors of $2-3$ at a single radius in the midplane, but are considerably thicker than AGN disks and neglect radiation altogether. These distributions correspond to $\sigma_T=0.1-0.2$. It is unclear whether the MRI-driven fluctuations in radiation-dominated global disks will be larger.

Thin, radiation-dominated disks have long been believed to be subject to a strong thermal instability, since the heating rate depends on a higher power of the temperature than the cooling rate.  \citet{hirosestable2009} find that the dynamics of shearing box simulations of radiation-dominated disks are well described by the alpha model on long timescales. The flow is highly variable, but thermally stable. Radiation-dominated disks are also subject to an inflow instability \citep{lightmaneardley1974}. Radially extended, radiation-dominated MHD simulations are needed to determine whether this instability operates in AGN. If so, it would likely create large local temperature gradients in the accretion flow between hot, tenuous gas and cooler optically thick regions. Finally, it is possible that a photon bubble instability could be present in radiation-dominated accretion disks \citep{turneretal2005}, which can cause the flux escaping the disk to rise by factors of several locally. This may prevent inhomogeneous disks from becoming dynamically unstable, despite exceeding the Eddington limit locally \citep{begelman2006}. The disk would also remain stable if magnetic pressure provides the vertical support in quasar accretion disks \citep[e.g.][]{begelmanpringle2007}. Alternatively, the disk may drive a wind when the Eddington limit is exceeded.

\section{Conclusions}

Standard thin disk accretion has formed the basis for understanding X-ray binaries and AGN for nearly 40 years. However, it has always been difficult to extend the model to explain optical quasar variability. In recent years, microlensing observations of multiply imaged quasars have provided a probe of the disk structure, finding that quasar microlensing sizes are robustly larger than the flux sizes predicted from thin disk theory. If the average  temperature structure remains identical to that in thin disk theory but is highly inhomogeneous, accretion disks can be large enough to explain the sizes found by microlensing while matching the observed level of optical variability. The level of inhomogeneity ($\sigma_T=0.35-0.50$ with $n=10^{2-3}$) required to explain the discrepancy in microlensing sizes is in excellent agreement with that necessary to fit observed quasar spectra. Such inhomogeneous structure produces short timescale variations in microlensing light curves that should be larger than $\simeq 1\%$. The amplitude of the temperature fluctuations can be further constrained from measuring the size of these deviations.  The range of temperatures in small regions of the inhomogeneous disk explains the simultaneous variability observed across optical wavelengths.

We have demonstrated this idea with an unphysical toy model. Proper modeling of an inhomogeneous disk will require global MHD simulations of radiation-dominated accretion disks. It is unclear whether the MRI alone is sufficient to produce the required temperature fluctuations, or whether additional disk instabilities or other variability mechanisms are also important.

\begin{acknowledgements}
We thank Robert Antonucci, Brandon Kelly, Joachim Wambsganss and the referee, Chris Kochanek, for useful comments. This work was partially supported by NASA Earth \& Space Science Fellowship NNX08AX59H. Support for program HST-GO-11225.02-A was provided by
NASA through a grant from the Space Telescope Science Institute,
which is operated by the Association of Universities for Research
in Astronomy, Inc., under NASA contract NAS 5-26555.
\end{acknowledgements}

\bibliographystyle{apj}

\end{document}